\documentclass[english,aps,prper,reprint,showpacs,titlepage,longbibliography]{revtex4-2}   

\usepackage[T1]{fontenc}	
\usepackage[latin9]{inputenc}	
\usepackage{geometry}		
\usepackage{graphicx}
\usepackage[above,below]{placeins}	
\usepackage{times}

\usepackage{hyperref}  
\hypersetup{colorlinks=true,urlcolor=blue,citecolor=blue,linkcolor=blue}   
\usepackage{enumitem} 
\setlist{nosep} 

\usepackage{outlines}
\usepackage{enumitem}
\setenumerate[1]{label=\Roman*.}
\setenumerate[2]{label=\Alph*.}
\setenumerate[3]{label=\roman*.}
\setenumerate[4]{label=\alph*.}

\begin{document}

\begin{titlepage}

  \title{Towards understanding and characterizing expert covariational reasoning in physics}

  \author{Charlotte Zimmerman}
  \author{Alexis Olsho}
  \author{Suzanne White Brahmia}
    \affiliation{Department of Physics, University of Washington, Seattle, WA, 98195}
  \author{Michael Loverude}
  \affiliation{Department of Physics, California State University Fullerton, 800 N. State College Blvd., Fullerton, CA, 92831} 
  \author{Andrew Boudreaux}
  \affiliation{Department of Physics, Western Washington University, 516 High Street, Bellingham, WA,
    98225}
  \author{Trevor Smith}
  \affiliation{Department of Physics and Astronomy and Department of STEAM Education, Rowan University, 201 Mullica Hill Road, Glassboro, NJ,
    08028}


  \begin{abstract}
    Relating two quantities to describe a physical system or process is at the heart of ``doing physics'' for novices and experts alike. In this paper, we explore the ways in which experts use covariational reasoning when solving introductory physics graphing problems. Here, graduate students are considered experts for the introductory level material, as they often take the role of instructor at large research universities. Drawing on work from Research in Undergraduate Mathematics Education (RUME), we replicated a study of mathematics experts' covariational reasoning done by Hobson and Moore with physics experts [N. L. F. Hobson and K. C. Moore, in \emph{RUME Conference Proceedings}, pp. 664-672 (2017)]. We conducted think-aloud interviews with 10 physics graduate students using tasks minimally adapted from the mathematics study. Adaptations were made solely for the purpose of participant understanding of the question, and validated by preliminary interviews. Preliminary findings suggest physics experts approach covariational reasoning problems significantly differently than mathematics experts. In particular, two behaviors are identified in the reasoning of expert physicists that were not seen in the mathematics study. We introduce these two behaviors, which we call \textit{Using Compiled Relationships} and \textit{Neighborhood Analysis}, and articulate their differences from the behaviors articulated by Hobson and Moore. Finally, we share implications for instruction and questions for further research.  \clearpage
  \end{abstract}

  \maketitle
\end{titlepage}

\section{Introduction} 
Covariational reasoning---how a change in one quantity results in a change in another, related quantity---is integral to how scientists model and understand the physical world. While physics education researchers have long been considering the role of mathematics in physics, the details of how physics experts and novices coordinate changes in two related quantities is not yet well understood. In this work, we replicate a study of mathematics graduate students' covariational reasoning with physics graduate students to explore how experts in physics reason covariationally  \cite{Hobson2017ExploringReasoning}.

Notable research in physics education has examined the role of conceptual mathematics in how students reason about physics \cite{Uhden2012ModellingEducation, Karam2014FramingElectromagnetism, VonKorff2014DistinguishingPhysics, Caballero2015UnpackingHere, Ibrahim2017HowPerformance}. In particular it is evident that the use of mathematics is context driven and therefore fundamentally different in a physics setting than a purely mathematical one \cite{Bing2006TheKnowledge, Kuo2012HowProblems, Brahmia2016ObstaclesPhysics, Redish2015LanguageEpistemology}. Recent work by White Brahmia and collaborators defines \textit{physics quantitative literacy} (PQL) as ``the rich ways that physics experts blend conceptual and procedural mathematics to formulate and apply quantitative models'' \cite{Olsho2019TheSign}. In developing an inventory to assess quantitative literacy in physics, Smith et al. identified three specific elements of PQL: proportional reasoning, reasoning about sign, and covariational reasoning \cite{Olsho2019TheSign, Smith2019DevelopingLiteracy, Brahmia2019APhysics}. We aim to further explore covariational reasoning as it is ubiquitous in the physical sciences and necessary to understand and conceptualize rates of change \cite{Thompson1994ImagesCalculus}.

Although little has been published about covariational reasoning in introductory physics contexts, significant work has been done to characterize students' covariational reasoning by the Research in Undergraduate Mathematics Education (RUME) community \cite{Johnson2015TogetherChange, Moore2013CovariationalSystems, Castillo-Garsow2013, Paoletti2017TheReasoning, Carlson1998AConcept}. It is an essential tool to interpret derivatives and rates of change of quantities and an important mental habit for student success in conceptualizing calculus \cite{Thompson1994ImagesCalculus, Saldanha1998Re-thinkingVariation, Oehrtman2008FoundationalEds.}. A framework presented by Carlson et al. articulates a collection of mental actions that can be used to classify types of covariational reasoning \cite{Carlson2002ApplyingStudy}. Here, \emph{mental action} is used to describe a specific behavior in students' reasoning, as defined by Carlson et al. For example, one mental action that is indicative of covariational reasoning is identifying that an increase in one quantity is related to a decrease in another. After reviewing this work, we asked the question: would the mental actions articulated by Carlson et al. be the same for \emph{physicists} engaged in covariational reasoning?

In 2017, mathematics education researchers Hobson and Moore published a study in which they interviewed graduate students in mathematics to characterize expert thinking related to covariational reasoning in graphical contexts \cite{Hobson2017ExploringReasoning}. Here, \emph{expert} refers to students enrolled in a PhD program with significant experience with the introductory-level material presented in the study. We sought to investigate the extent to which the behaviors identified in their work were present in physics student reasoning. Therefore, we replicated the Hobson and Moore study through similarly formatted think-out-loud interviews with physics graduate students. We then analyzed their apparent mental actions and reasoning patterns in order to determine how covariational reasoning compares between experts in mathematics and physics. 

Our aim in this study is to explore the differences in covariational reasoning between experts in mathematics and physics. In doing so, we found that physics expert reasoning exhibited novel behaviors that were not reported in previous work. We offer suggestions for instruction and further study. 

\section{Review of Prior RUME Study} 
\label{section:RUME}
\begin{figure}[h!]
    \centering
    \includegraphics[width=0.45\textwidth]{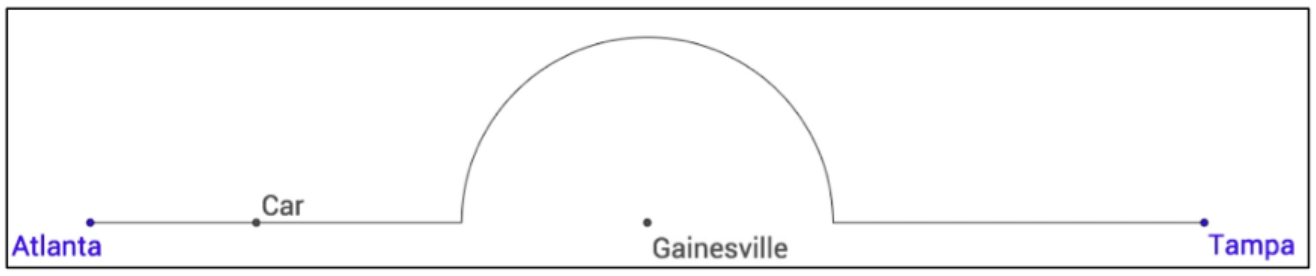}
    \caption{The figure presented by Hobson and Moore for the Going Around Gainesville task. \cite{Hobson2017ExploringReasoning}}
    \label{fig:gainesville}
\end{figure}

Hobson and Moore report conducting 10 think-out-loud interviews in which mathematics graduate students were given three tasks designed to elicit covariational reasoning. In each task, participants viewed an animation of motion and were asked to produce a graph to show the relationship between two specified distances. The first task involves a road trip from Atlanta to Tampa and asks interview participants to ``create a graph that relates their total distance traveled and their distance from Gainesville during the trip'' (Fig.~\ref{fig:gainesville}) \cite{Hobson2017ExploringReasoning}. The second task shows a cart going around a Ferris wheel and asks students to relate the height of the cart to the total distance traveled; the third task depicts a car moving around a square track and asks students to relate the distance of the car to the wall to the total distance traveled. The researchers employed grounded theory and conceptual analysis to analyze their results, and present the reasoning patterns of two students in particular that are representative of their findings.

Hobson and Moore identified several reasoning behaviors, but we focus on three mathematics expert behaviors, or MEB: 
\begin{enumerate}
    \item Coordinating amounts of change of two quantities, that is, explicitly recognizing that a change in one quantity was directly related to a change in the other.
    \item Dividing the representation into equal sections to examine how a quantity changes during equal changes of another quantity and comparing the rates of change across sections.
    \item Explicitly referring to the changing rate of change in a quantity and its relationship to the curvature of the graph.
\end{enumerate}
When considering coordinated amounts of change, students may, for example, recognize that a change in the total distance traveled in the Gainesville task is connected to a corresponding change in the distance to Gainesville. Hobson and Moore report one student saying: ``So, whatever it is doing distance-wise is also happening to its distance to Gainesville\ldots10 miles total distance traveled and it's also 10 miles closer to Gainsville'' (MEB I). In the Ferris wheel task, Hobson and Moore report that students equally divided the Ferris wheel into wedges of equal arc length or equal changes of horizontal motion. Students then compared sections to see how the rate of change of the height was changing (MEB II). Finally, throughout the tasks and students presented, there are explicit references to the way that the rate of change is changing as a quantity is varied (MEB III) \cite{Hobson2017ExploringReasoning}. 

\section{Methods}

Our replication involved 10 individual interviews with physics PhD students at a large research university in the Northwest United States. We attempted to recruit a pool of participants as diverse in race, gender, experience level, and disciplinary focus as was possible within the limits of the population at the University. Interviews lasted between 25 and 50 minutes, and participants were compensated with \$15 gift cards to a local coffee shop. Each interview subject was given a computer with the animations and a question sheet. They had the freedom to play the short animations (approximately 5 seconds in length) at any time during the interview, and move between questions at will. 

\begin{figure}[h!]
    \centering
    \includegraphics[width=0.45\textwidth]{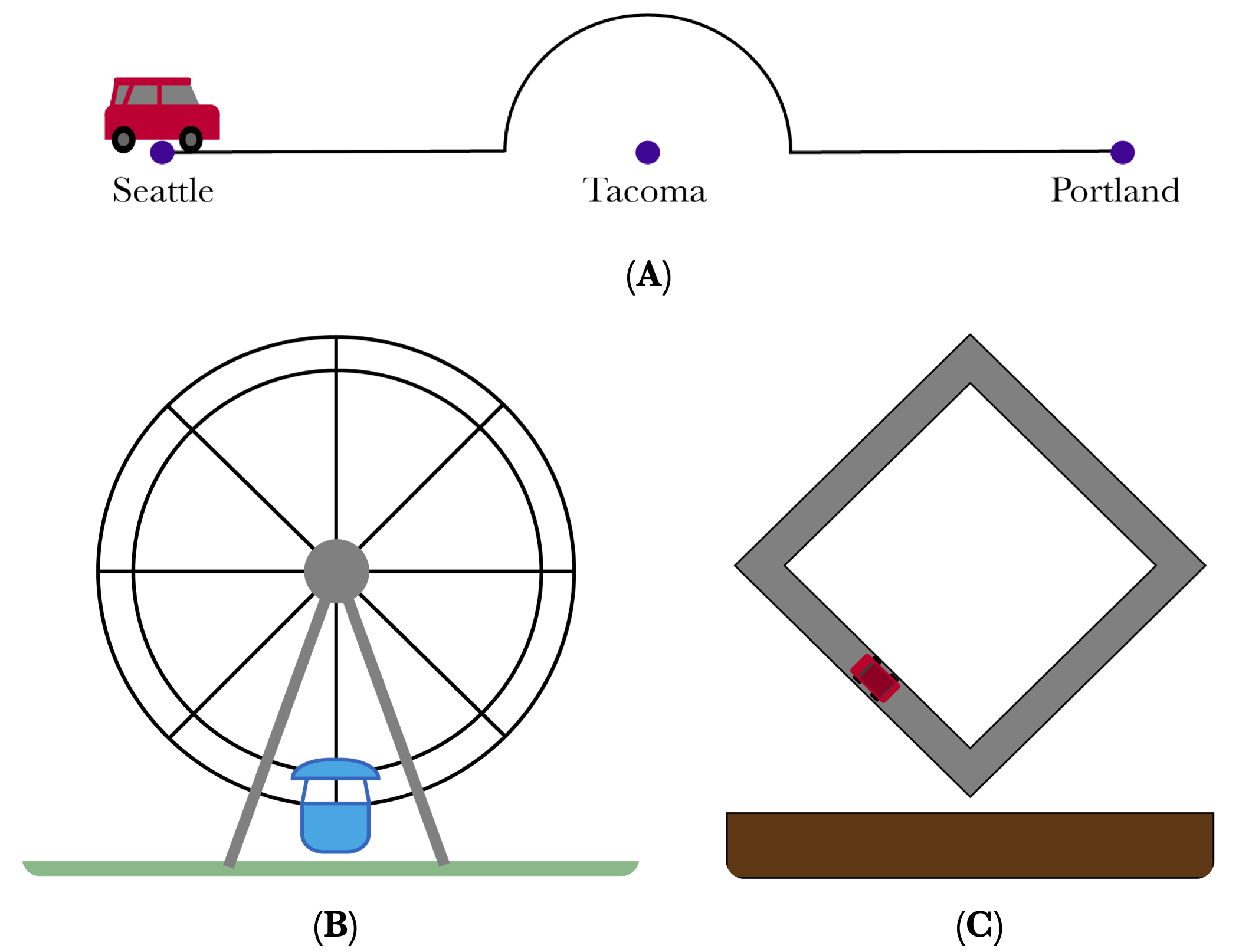}
    \label{fig:reasoningTasks}
    \caption{Interview tasks given to physics graduate students. (a) Going Around Tacoma: students were given the instructions, ``A few students are going on a road trip, and decide to drive from Seattle to Portland. They want to avoid the traffic around Tacoma, and take the path shown in the animation. Draw a graph of the students' distance from Tacoma vs. their total distance traveled in the space below.'' (b) Ferris Wheel: students were given the instructions, ``As shown in the animation, a cart moves around a Ferris wheel. Draw the graph of the height of the cart from the ground vs its total distance traveled in the space below.'' (c) Square Track: students were given the instructions, ``As shown in the animation, a car moves around a track. Draw the graph of the distance of the car from the wall vs its total distance traveled in the space below.''}
\end{figure}

As the goal of the interviews was a direct replication of Hobson and Moore's study, care was taken to mimic the animations and text of the interview questions used by Hobson and Moore. However, small changes in language were made for clarity purposes. The animations shown in Fig~\ref{fig:reasoningTasks} were developed to be functionally identical to those used in the RUME study, based on the descriptions and stills presented there (see Sec~\ref{section:RUME} for a description). Although each animation depicts a constant speed motion, participants were not explicitly told that the speeds were constant. 

Initial versions of transcripts were generated from recorded student audio with the computer program Otter, and subsequently hand-corrected \cite{2019Otter}. Participants' written work was also collected and examined. Transcripts were analyzed using Process and In Vivo coding to identify patterns in reasoning \cite{Saldana2016TheResearchers}. Our coding was informed by the Hobson and Moore study, and by our literature review on covariational reasoning.  As this review was concurrent with our coding, the analysis is best described as a modified approach to grounded theory \cite{Glaser1967TheTheory}.

\section{Results}
In this paper, we aim to present preliminary findings based on ongoing analysis of the interview transcripts. In our preliminary coding scheme, we identified eight categories of reasoning that were being used by physics experts. Some of these are consistent with the reasoning shown by Hobson and Moore, and some are not. Similarly, some were coded across all participants and some were only coded for a few. Therefore, we focus here on two behaviors that all participants demonstrated at least one of in their reasoning, and that illustrate the similarities and differences between expert physics and mathematics reasoning. 

In the first reasoning mode, which we refer to as ``Using Compiled Relationships,'' participants deploy a relationship between familiar quantities that they have produced or have prior knowledge of to coordinate changes in the two covarying quantities. In the second, ``Neighborhood Analysis,'' participants choose a specific point or set of points that are especially relevant, determine the behavior of one quantity with respect to another at those points, and then connect the slopes between these points to complete the graph. Throughout our analysis, participants were all coded as using one of these two categories when attempting to explain their reasoning. To better demonstrate how these reasoning modes manifest, we will focus on one  transcript in some detail: an interview with a 4th year physics PhD student whom we call ``Oliver''. We chose Oliver as an example of our findings as they use both approaches and clearly articulate their reasoning during the interview.

\subsection{Using Compiled Relationships}
In this mode of reasoning, participants identified an alternative quantity based on a familiar model rather than considering the covariational relationship directly. For example, many participants substituted time for the total distance traveled, or applied familiar trigonometric functions to the uniform circular motion of the Ferris wheel. Both of these reasoning moves were nearly universal in the interviews conducted, with some exceptions discussed below. We speculate that physics experts may be reducing cognitive load by eliminating the need to directly determine the relationship between the two given quantities. Instead, the familiar models of uniform motion and sinusoidal variation allow the experts to use a previously well understood covariational relationship to explain the new situation.

\begin{figure}
    \centering
    \includegraphics[width=0.43\textwidth]{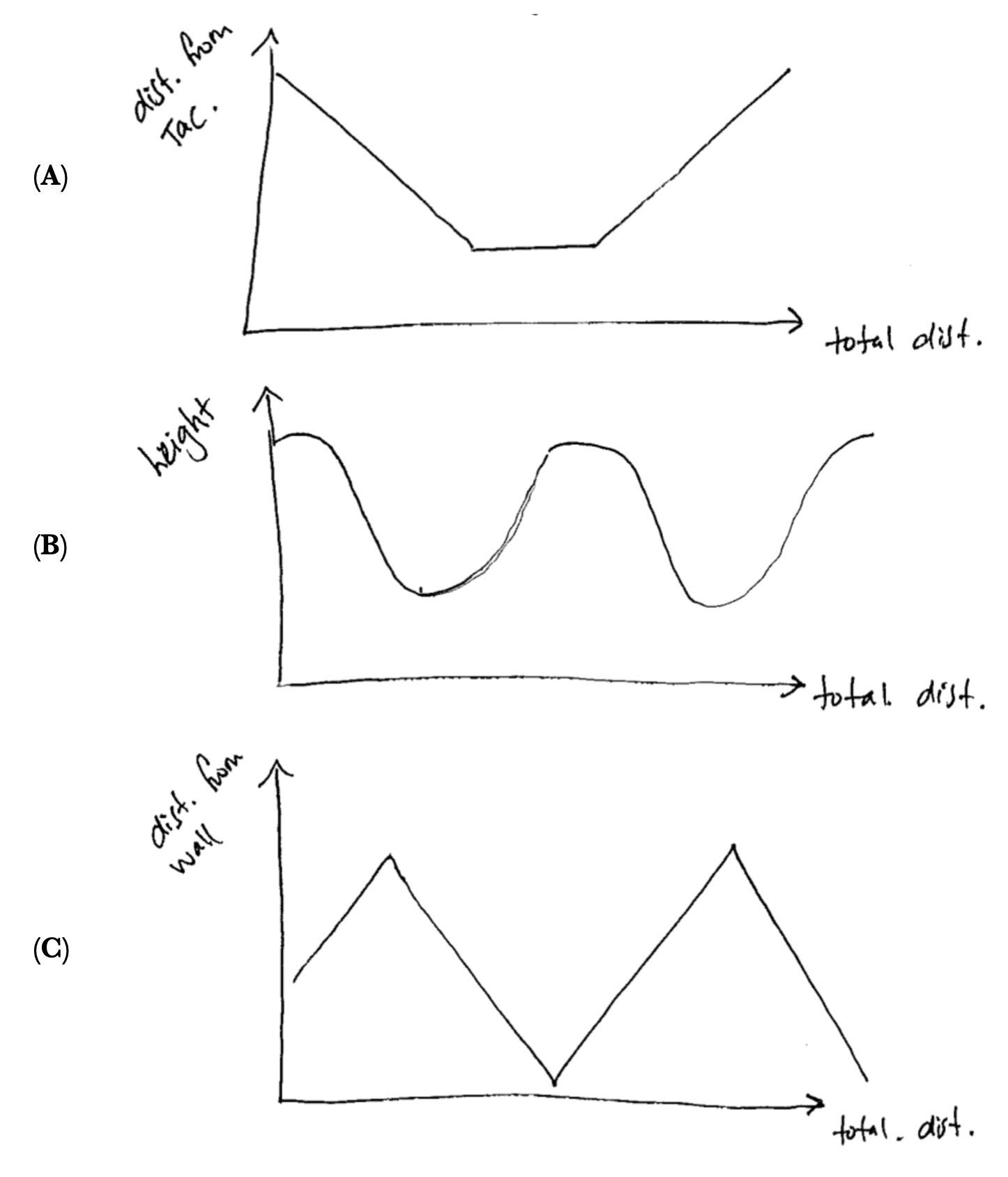}
    \caption{The graphs generated by Oliver during the course of the interview. We consider these graphs to be correct answers to the tasks at hand. (A) corresponds to the Going Around Tacoma task, (B) to Ferris Wheel, and (C) to Square Track.}
    \label{fig:graphs}
\end{figure}

On the Going Around Tacoma task, Oliver began by considering what quantities to use on their axes: ``So distance from Tacoma is on the Y-axis, and then total distance\ldots it's going to be on the X-axis.'' At this point, Oliver referred back to the animation, noted that the car was moving at constant speed, and concluded that they could use time to draw their graph: ``It looks like they're going at a constant speed. So, total distance is just going to map to time\ldots I can think about that being time.'' Oliver's focus on a salient feature of the car's motion, constant speed, allows them to ``map'' the total distance traveled onto the time elapsed. They recognize that replacing total distance with time will yield the same graph. We interpret this as Oliver applying a quantitative model of constant speed, $x = vt$. Since graphing distance versus time is a well-practiced skill for expert physicists, Oliver has reduced the problem to something familiar and is able to quickly draw a correct graph (see Fig.~\ref{fig:graphs}A). This approach differs notably from the reasoning of the mathematics graduate students, who tended to develop direct relationships between distance from Gainesville and total distance traveled by coordinating changes between those two quantities as the car moved along its path (MEB I) \cite{Hobson2017ExploringReasoning}.

Similarly, in the Ferris wheel task, Oliver uses prior knowledge of uniform circular motion to avoid directly analyzing the coordinated changes between the height and the total distance traveled. Upon reading the problem statement, Oliver says: ``I feel like this is where, like, my understanding of trig functions really comes in handy. Because I know this is a circle. And so, the height goes like a trig function.'' The relationship between trigonometric functions and circles is strong enough in an expert physicist to immediately trigger this response, and Oliver identifies it themselves. This again differs from the sectioning behavior that the mathematics graduate students engaged in (MEB II). Indeed, even when asked to explain in more detail how they were thinking about the relationship between height and total distance, Oliver simply states: ``It's just something that's moving around in a circle\ldots I'm just like---oh, trig function.'' Only after significant probing did Oliver produce an explanation unambiguously recognizable as covariational reasoning---explicitly discussing how and why the height changes with respect to total distance traveled as the cart moves around the circle. At that point, Oliver used Neighborhood Analysis to explain why a trigonometric function applies to the Ferris wheel task, as discussed in the next section. Not every physics expert in our study readily identified the trigonometric behavior in the Ferris wheel task. Those who did not use a trigonometric model engaged directly with Neighborhood Analysis.

\subsection{Neighborhood Analysis}
In Neighborhood Analysis, the participant identifies specific points and determines the slope near those points to find the overall behavior of the graph. We observed this behavior throughout our analysis, but focus on three examples here: using the ends of the semi-circle to begin the graph in the Going Around Tacoma task, using the corners of the square to draw the graph in the Square Track task, and using the top, bottom, and side points of the Ferris wheel to explain and derive trigonometric behavior. Neighborhood analysis contrasts with the expert mathematicians' consideration of changing rate of change (MEB III). Expert physicists avoid explicit analysis of covariation \emph{between} important points by focusing on the rate of change only at the points themselves. We speculate that the physics experts were again reducing cognitive load, in this case by avoiding consideration of a compound (i.e. second order) rate of change.

In the first problem, after Oliver has decided to consider the graph of distance from Tacoma versus time, they note that the circular path around Tacoma corresponds to a straight line on the graph. After drawing a straight line in the middle of their graph, Oliver states: ``and then it's going to be increasing linearly on either side of that,'' suggesting implicit use of the points at either end of the semi-circular path to partition their graph into sections and consider the behavior in each. This behavior occurs again during Oliver's work on the Square Track task. Oliver begins by drawing axes and applying the time-for-total-distance model, as before. In this case, however, they are more explicit about their choice of points: ``So at the middle, it's going to be closest to the wall, and then the distance from the wall doesn't depend on that turn it takes on the sides. So its just going to be something like this [draws graph shown in Fig.~\ref{fig:graphs}C].'' Oliver chooses physically relevant points (in this case, associated with a change in direction of the car), considers how the relevant quantity is changing at those points, and draws their graph accordingly. 

Neither the Going Around Tacoma task nor the Square Track task allowed us to probe the extent to which participants engage in direct analysis of a changing rate of change, since they both involve linear relationships. To explore this, we examined how Oliver explained their reasoning in the Ferris wheel task after probing from the interviewer. Recall that Oliver had readily applied a trigonometric function to the Ferris wheel task and produced their graph using that model. After questioning from the interviewer, Oliver explains why a trigonometric model made sense: ``At the very start, and then down at the bottom\ldots if you take a small chunk of the height part, it's not changing very much. In the middle here\ldots when it's on the side of the Ferris wheel\ldots that's when the height's changing a bunch.'' Oliver identifies important points along the trajectory (the top, bottom, and sides), and describes how the height changes with small movement along the trajectory around those points. They consider this to be a sufficient basis for generating the graph. These observations are similar to the sense-making work done by Lenz et al. \cite{Lenz2018StudentStudy}.

We observed a similar approach to generating the graph among participants who did not immediately note the sinusoidal relationship between height and distance traveled. For example, one participant begins the Ferris wheel task by identifying the same points on the trajectory (top, bottom, and sides) and discussing how the height changes around those points. They eventually conclude that the change should be slow on the top, fast on the sides, and slow on the bottom. The participant explains: ``So if I want it to be slow, fast, slow\ldots I think it looks like\ldots [draws a sinusoidal-shaped curve].'' The participant identifies physically significant points, and analyzes the slopes at those points to produce a graph. This behavior seems to additionally reduce cognitive load, and is notably distinct from the process of directly analyzing the change in the rate of change of a quantity (MEB III).

\section{Conclusion}
In our replication study, we found two behaviors in preliminary analysis that were consistent across expert physicists and distinct from those of mathematics experts. We observed that physics graduate students sought alternatives to directly analyzing the relationship between two distances, and that they approached producing a relationship by examining change at ``important'' points rather than assessing the rates of change of equally spaced sections. These differences suggest that physics experts employ a different set of covariational reasoning skills than mathematicians. 

We do not propose that our findings are the only ways that covariation emerges in physics reasoning. Indeed, it is likely that our preliminary investigation into covariation in these graphing contexts is merely a small facet of covariational reasoning in physics. Therefore, we suggest that further research into the covariational reasoning of physics experts is warranted. Such research would aid in the development of effective teaching strategies for the modes of covariational reasoning necessary in introductory-level physics. Our findings suggest that students starting an introductory physics course might not have been exposed to these physics-specific reasoning skills in prior math classes.

We are continuing our work to determine if the patterns observed in introductory-physics tasks continue with more advanced physics-content questions. This includes tasks for which participants cannot easily substitute time for one of the variables, or readily apply a well-practiced model. We also suggest that instructors consider explicit use and discussion of covariational reasoning skills and strategies in their teaching. Explicit identification of such skills during instruction, especially those skills that may not be taught in math courses, may help students build their own facility. In particular, instructors may find it useful to discuss how to identify important points in a problem and when it may be productive to simply ``connect slopes'' rather than consider the behavior of the function between two points. 

We claim that covariational reasoning manifests in some different ways in physics than it does in mathematics. We believe this is likely due to differences in the role and goals of mathematics in the two disciplines. We suggest that covariational reasoning, which is at the heart of sense-making with physics models, be further explored through research in physics education. A better understanding and teaching of these skills is of central importance for students continuing in physics as well as in other STEM disciplines.

\acknowledgments
We would like to thank Natalie Hobson and Kevin C. Moore for their interest and permission for our replication of their work with physics graduate students. This work was made possible by funding from the University of Washington and by the National Science Foundation IUSE grants: \textit{DUE-1832880, DUE-1832836,} and \textit{DUE-1833050.}

\bibliographystyle{apsrev}
\bibliography{Towards_Understanding_and_Characterizing_Expert_Covar} 

\end{document}